# Microwave Generation Power in a Nonrelativistic Electron Beam with Virtual Cathode in a Retarding Electric Field

E. N. Egorov, Yu. A. Kalinin, A. A. Koronovskiĭ, A. E. Hramov*, and M. Yu. Morozov

*Saratov State University, Saratov, Russia*
*e-mail: aeh@nonlin.sgu.ru*

**Abstract**—The power of microwave generation in a nonrelativistic electron beam with virtual cathode formed in a static retarding electric field (low-voltage vircator system) has been studied experimentally and by means of numerical simulation within the framework of a one-dimensional theory. The limits of applicability of the one-dimensional theory have been experimentally determined.

Electron-wave systems with the active medium in the form of an electron beam with a virtual cathode (VC) are among the promising sources of high-power microwave radiation [1–4] and are extensively studied in the context of various applications [5, 6].

Recently [7–9], we proposed and studied a new scheme of the oscillator with a VC (vircator) making use of an intense nonrelativistic electron beam (i.e., a beam with a microperveance $p_\mu > 3$ µA/V$^{3/2}$ [10]). In order to form a VC in the electron beam, this system employs a scheme with an additional retardation of electrons. According to this scheme (called a low-voltage vircator), a nonstationary oscillating VC is formed at the expense of strong retardation of electrons in the drift region, which makes possible the generation of both single-frequency and broadband microwave signals using electron beams with small total currents and low densities [7]. In such regimes, it is possible to study in much detail the physical processes in electron beams with VCs. It should be noted that the system with a VC and a retarded electron beam is also of interest as a controlled source of medium-power broadband chaotic signals in the centimeter and millimeter wavelength range [8].

This Letter presents the results of an experimental and theoretical investigation of the output energy characteristics of a low-voltage vircator.

The experimental investigation of oscillations in the electron beam with a VC was performed using as diode scheme in which the beam formed by an electron-optical system (EOS) was injected into a system (see Fig. 1) comprising two grid electrodes (5 and 6) forming a retarding field. The retarding field was created by applying a negative potential $V_r$ to the exit (second) grid 6 relative to the entrance (first) grid 5. The EOS formed an axisymmetric converging cylindrical electron beam 4. The accelerating voltage in our experiments was 2.0 kV, the beam current at the EOS output was varied within 50–100 mA (depending on the cathode filament voltage), the EOS length was $l = 70$ mm, and the beam radius was $r_b = 4$ mm.

The electron beam generated in the EOS with a certain initial scatter of electron velocities enters into the space between grids (diode gap). The first grid poten-tial $V_0$ is equal to the anode potential (i.e., to the accelerating voltage); the second grid potential $V_r = V_0 - \Delta V_r$ varies from $V_r/V_0 = 1$ ($\Delta V_r = 0$, no retardation) to $V_r/V_0 = 0$ ($\Delta V_r = V_0$, complete retardation). As the retarding potential difference $\Delta V_r$ between grids of the diode gap is increased to reach a certain critical value $[\Delta V_r]_c$, an oscillating VC is formed [7] from which electrons are partly reflected back to the entrance grid. As a result, chaotic oscillations are generated, the shape and power of which are significantly dependent on the $\Delta V_r$ value.

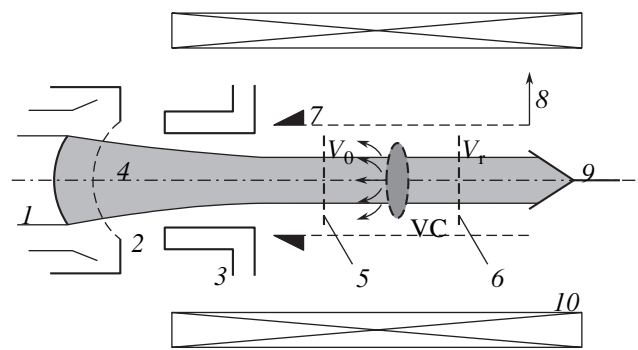

**Fig. 1.** Schematic diagram of the experimental setup used for the investigation of oscillations in a nonrelativistic electron beam with virtual cathode formed in a system with additional retardation of electrons: (*1*) thermionic cathode; (*2*) grid electrode of the electron gun; (*3*) second anode of the electron gun; (*4*) converging electron beam formed by the EOS; (*5*) entrance grid of the diode gap; (*6*) exit grid of the electron gun (to which the retarding potential $V_r$ is applied); (*7*) helical slow-wave system; (*8*) energy output; (*9*) collector; (*10*) solenoid.

The power of VC oscillations is extracted via a broadband helical slow-wave system 7 (HSWS) loaded on an absorbing output insert 8. The HSWS coupling resistance is $K = 100\ \Omega$ and the retardation factor is $n \approx 10$. Past the diode gap, the electron beam strikes collector 9 (Fig. 1). The experiments were performed in a continuously evacuated vacuum setup with a minimum residual pressure of $10^{-7}$ Torr.

Figure 2a (curve 1) shows an experimental plot of the total oscillation power $P$ versus relative retarding potential difference $\Delta V_r/V_0$ between grids of the diode gap. The electron beam was focused by a homogeneous guiding electric field ($B = 200$ G) generated by solenoid 10 (Fig. 1). For $\Delta V_r/V_0 < 0.2$, the electron beam exhibits no oscillations ($P = 0$). As the electron beam retardation is increased, a VC is formed in the beam. However, as can be seen from Fig. 2a, the total power of oscillations in the system with a VC is initially small. As the $\Delta V/V_0$ value is increased further, the output power also grows, reaches a maximum value of $P \approx$ 150 mW for a certain optimum retarding potential, and then decreases again. It should be noted that VC oscillations observed in the regime of small retardation are close to regular, and the spectrum of microwave radiation is discrete. Upon a further increase in the relative retarding potential difference, broadband chaotic oscillations arise in the system [7], but a large retardation ($\Delta V_r/V_0 > 0.75$) leads to the complete suppression of generation in the system.

Nonlinear nonstationary processes in a beam of charged particles with a VC were simulated using the particle-in-cell method [11, 12] within the framework of a one-dimensional model of the drift region with a retarding field. The space-charge fields were determined by solving a one-dimensional Poisson equation. Thus, the motion of electrons in the beam was considered as one-dimensional; this assumption is valid, but not in all regimes of the vircator system under consideration. The scheme of numerical calculations used in this study is described in detail elsewhere [7].

In the theoretical study of the microwave generation power, the output HSWS was modeled (following [13, 14]) by an equivalent long transmission line comprising serially connected inductances $L$ [H/m] shunted by capacitances $C$ [F/m]. Assuming a paraxial electron beam propagates along the long line so that all field lines corresponding to a separate parts $q(x)$ of the charge originate at a point with the same longitudinal coordinate $x$ and terminate on the line, the beam charge $q(x)$ can be considered as induced in the long line. Then, the HSWS can be described by telegrapher's equations with an additional term corresponding to the excitation of electromagnetic waves by the beam:

$$\frac{\partial I}{\partial t} = -\frac{1}{L}\frac{\partial U}{\partial x}, \quad \frac{\partial U}{\partial t} = -\frac{1}{C}\frac{\partial I}{\partial x} + \frac{1}{C}\frac{\partial q}{\partial t}. \quad (1)$$

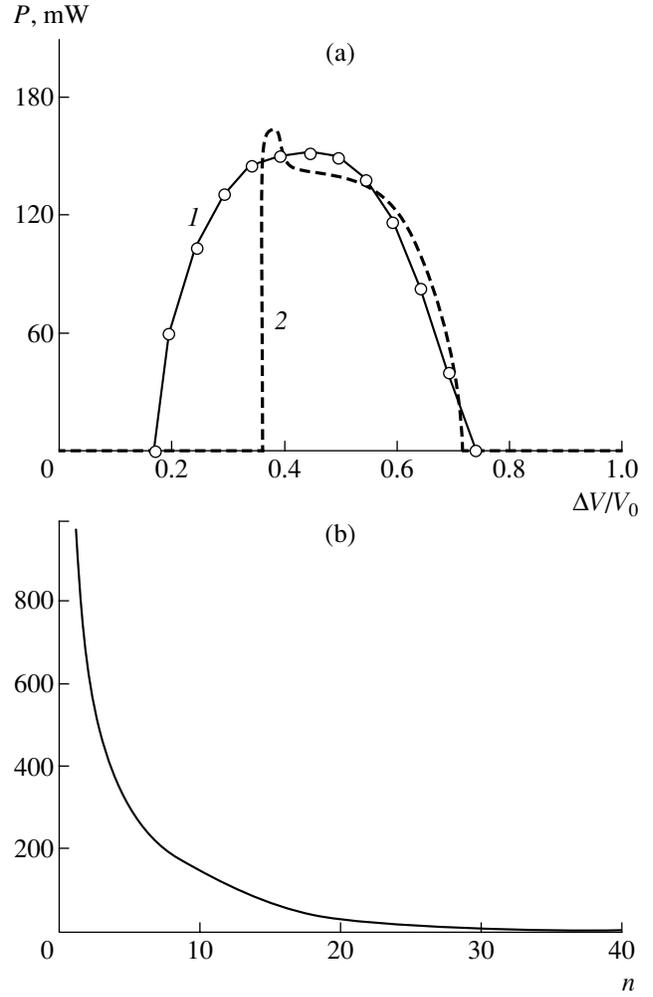

**Fig. 2.** Plots of the power $P$ of electromagnetic oscillations at the HSWS output (a) versus retarding potential ($\Delta V_r/V_0$ (1, experiment; 2, numerical calculations using formula (2) of a nonlinear nonstationary theory) and (b) versus wave retardation $n = v_{ph}/c$ in the transmission line at $\Delta V_r/V_0 = 0.5$.

We also assume that the reverse influence of the HSWS field on the charge dynamics can be ignored (preset current approximation). This assumption is valid provided that the electron flux is not synchronized with the electromagnetic wave propagating in the long line under consideration. Telegrapher's equations (1) were numerically solved assuming that the line is matched on both left ($x = 0$) and right ($x = l$) ends of the HSWS. The charge distribution $q(x, t)$ in the beam exciting electromagnetic waves in the transmission line was determined using a numerical solution of the problem (described in detail in [7, 9]).

Since oscillations in the beam with a VC are far from being harmonic, the output radiation power of the system under consideration can be expressed as

$$P = \frac{1}{TK}\int_0^T U^2(t, x = l)dt, \qquad (2)$$

where $l$ is the length of the transmission line and the integral implies time averaging of the instantaneous power $P_m(t) = U^2(t, l)/K$ at the right (exit) end of the transmission line.

Figure 2a (dashed curve 2) shows a plot of the output electromagnetic field power $P$ versus retarding potential $\Delta V_r/V_0$, which was also calculated within the framework of the nonlinear nonstationary theory. A comparison of the experimental and theoretical curves shows their good qualitative and quantitative agreement at large values of the retarding potential ($\Delta V_r/V_0 > 0.4$). In this regime, the nonstationary one-dimensional theory is valid, that is, the electron beam dynamics in a vircator with large values of the retarding potential and the focusing magnetic field is close to one-dimensional.

However, the situation changes at small values of the retarding potential $\Delta V_r/V_0$: the experimental curve in Fig. 2a shows that the power of generation gradually decreases in the region of $\Delta V_r/V_0 < 0.4$ and vanishes at $\Delta V_r/V_0 \approx 0.18$. Thus, oscillations in the beam with a VC softly arise with increasing retardation, and the generation region extends over the range of $\Delta V_r/V_0 \in [0.18, 0.72]$. In contrast, the results of numerical simulation are indicative of a sharp breakage of oscillations at $\Delta V_r/V_0 \approx 0.38$. Accordingly, the region of existence of the VC oscillations decreases, which is at variance with the experimental data.

The results of the above analysis suggest that the discrepancy between theory and experiment in the region of small retarding potentials is related to the neglect of two-dimensional motion of electrons in the interaction space. At small retarding potentials in the VC region, the beam dynamics becomes substantially two-dimensional and there is a significant current leakage to the HSWS, so that the one-dimensional model fails to be applicable under such conditions [7].

Now let us briefly consider the problem of determining the optimum parameters of the transmission line, for which the microwave radiation power extracted from the VC region reaches a maximum. On the one hand, an increase in the output power can be provided by the increase in the resistance $K$ of coupling between the beam and the transmission line. However, the possibilities of significantly increasing $K$ are restricted, and the experimental estimate ($K = 100\ \Omega$) is close to the maximum possible value. Another possibility is offered by a change in the phase velocity $v_{ph}$ of the wave in the transmission line. Figure 2b shows a plot of the output power versus wave retardation expressed as $n = v_{ph}/c$, where $c$ is the velocity of light. As can be seen from these data, the generation power $P$ grows with decreasing $n$ and reaches a level of $P \approx 0.6$–$0.8$ W at small $n$. This behavior is probably related to the fact that the VC oscillator, in contrast to the systems with long-term interaction (such as traveling-wave tubes and backward-wave oscillators), do not obey the condition of synchronism between the wave and the electron flow, according to which the phase velocity of the electromagnetic wave must be close to the velocity of electrons in the beam [14]. Moreover, this synchronism is in fact impossible in view of the significant change in the velocity of electrons in the retarding field: from $v_0$ (determined by the accelerating voltage $V_0 = 2$ kV) to zero in the region of electron stopping and reflection (near the VC). However, it should be noted that a decrease in the retardation unavoidably leads to a corresponding decrease in the coupling resistance [15]. Therefore, a certain optimum design of the energy output probably exists, for which the output radiation power is maximum.

**Acknowledgments.** The authors are grateful to Prof. D.I. Trubetskov for his interest in this study, fruitful discussions, and useful critical remarks.

This study was supported by the Russian Foundation for Basic Research (project nos. 05-02-16286 and 05-02-08030) and by the "Dynasty" Noncommercial Program Foundation.